# Anomalous Nernst effect beyond the magnetization scaling relation in the ferromagnetic Heusler compound Co$_2$MnGa

Satya N. Guin[1,*], Kaustuv Manna[1], Jonathan Noky[1], Sarah J. Watzman[2], Chenguang Fu[1], Nitesh Kumar[1], Walter Schnelle[1], Chandra Shekhar[1], Yan Sun[1], Johannes Gooth[1,*], Claudia Felser[1,*]

[1] *Max Planck Institute for Chemical Physics of Solids, 01187 Dresden, Germany*

[2] *Department of Mechanical and Aerospace Engineering, The Ohio State University, Columbus, OH 43210, USA*

*Correspondence to: satyanarayanguin@cpfs.mpg.de*

*Johannes.Gooth@cpfs.mpg.de*

*Claudia.Felser@cpfs.mpg.de*

**Applying a temperature gradient in a magnetic material generates a voltage that is perpendicular to both the heat flow and the magnetization.**[1,2] **This is the anomalous Nernst effect (ANE),**[3,4] **which was thought to be proportional to the value of the magnetization for a long time. However, more generally, the ANE has been predicted to originate from a net Berry curvature of all bands near the Fermi level ($E_F$).**[5,6] **Subsequently, a large anomalous Nernst thermopower ($S_{yx}^A$) has recently been observed in topological materials with no net magnetization but large net Berry curvature [$\Omega_n(k)$] around $E_F$.**[7–9] **These experiments clearly fall outside the scope of the conventional magnetization-model of the ANE, but a significant question remains: Can the value of the ANE in topological ferromagnets exceed the highest values observed in conventional ferromagnets? Here, we report a remarkably high $S_{yx}^A$-value of ~6.0 μV K$^{-1}$ at 1 T in the ferromagnetic topological Heusler compound Co$_2$MnGa at room temperature, which is around 7-times larger than any anomalous Nernst thermopower value ever reported for a conventional ferromagnet. Combined electrical, thermoelectric and first-principles calculations reveal that this high value of the ANE arises from a large net Berry curvature near the Fermi level associated with nodal lines and Weyl points.**

Conventional thermoelectric devices are based on the Seebeck effect in a two-terminal geometry, where two electronic reservoirs at different temperatures result in an electric voltage that arises in the direction of the imposed temperature gradient.[1,2] For applications, however, this configuration has considerable drawbacks, as the heat reservoir is forced to be part of the electrical circuit. The problem is to achieve good thermal insulation while maintaining good electrical conduction. To circumvent this issue, multi-terminal thermoelectric devices have recently attracted increasing attention,[10–13] because they allow for the spatial separation of the heat reservoir from the electric circuitry. One possible implementation of this kind[13] is based on the Nernst effect, which is the generation of a transverse thermoelectric voltage by a temperature gradient and a magnetic field, applied normal to each other.[14–16] Nernst devices are, in principle, also simpler to integrate technologically than conventional thermoelectrics as there is no need for both *p*- and *n*-type materials.

A ferromagnetic material generates an anomalous transverse voltage mutually perpendicular to both heat current and magnetization is so-called anomalous Nernst effect (ANE).[3,4,17–20] The ANE is the thermoelectric counterpart of the anomalous Hall effect (AHE) and was first observed in topologically trivial ferromagnets. Therefore, it was believed for a long time that the ANE is proportional to the magnetization of a material itself. In spite of many efforts, the maximum value of anomalous Nernst thermopower in ferromagnetic systems is limited and bound to below ~1 µV K$^{-1}$.[9] However, recently it was found that the AHE is not directly related to the magnetization of a material but derived more generally from the Berry curvature [$\Omega_n(k)$] of the bands near $E_F$.[5,6] This was realized hand in hand with the discovery that the AHE is linked to the summation of $\Omega_n(k)$ over all the occupied bands below $E_F$.[21]

$\Omega_n(k)$ is a local gauge field associated with the Berry phase of the band structure and determines the topological aspects of a material. In contrast to previous believes it is thus possible to control the Berry curvature and thus the intrinsic AHE and ANE by suitable manipulations of symmetries and band structures, independent of the finite value of the magnetization. As a result of these considerations, the AHE and ANE have recently been tuned in a distinct manner by tailoring the distribution of $\Omega_n(k)$ in topological matter, such as in Dirac metals or chiral anti-ferromagnets.[5–9,22–25] These investigations have led to the observation of an AHE and an ANE in absence of ferromagnetism. Furthermore, is has been demonstrated that by manipulating the crystal and magnetic space group symmetry and details of band structure, the anomalous Hall conductivity can be tuned from zero to a large value.[26] The Berry curvature-driven anomalous Hall and Nernst effects have been measured in the non-collinear antiferromagnets $Mn_3Sn$ and $Mn_3Ge$, without sizeable magnetization, where Berry curvature effects are generated from the non-collinear spin structure.[9,22–25] Also, in the Dirac semimetal $Cd_3As_2$ the AHE and ANE were measured due to Berry curvature effects, produced by the separation of the Weyl nodes in magnetic field.[7,8] The results of both experiments clearly fall outside the scope of the conventional model which scales the ANE with the magnetization of the system, but a worthy question remains: Is it possible to push the limit of ANE beyond the magnetization-scaling relation in a ferromagnet? Hence, this work desire to go beyond the recent experiments, identify a material with a large $\Omega_n(k)$ at $E_F$ and perform Nernst measurements on this system.

In the present study, we have carried out thermoelectric experiments on the ferromagnetic topological Heusler compound $Co_2MnGa$, which has previously been shown to exhibit a large AHE and a large Berry curvature-distribution around $E_F$.[26,27] $Co_2MnGa$ is a full Heusler compound, which crystallizes in the L2$_1$ structure with space group $Fm\bar{3}m$ and a lattice constant $a = 5.77$ Å (Fig.1 (a)). The structure consists of four interpenetrating fcc sub-lattices, of which two are formed by Co atoms. The other two sub-lattices are formed by the Mn and Ga atoms. The structure can also be implicitly viewed as a zinc-blende-type sub-lattice, formed by one Co and Ga in which tetrahedral and octahedral holes are occupied by the second Co and Mn atom respectively. The Co atoms occupy the Wyckoff position 8c (1/4,1/4,1/4), whereas the Mn and Ga atoms occupy the Wyckoff positions 4a (0, 0, 0) and 4b (1/2,1/2,1/2), respectively. Without spin-orbit coupling, there are three nodal lines near $E_F$ (Fig. 1 (b) and (c)). The spin-orbit coupling lifts the degeneracy in two of these lines and a small gap is developed. This effect induces a large $\Omega_n(k)$ into the system at the former nodal lines, making the material a promising candidate for the study of Berry curvature related effects. A recent angle-resolved photoemission spectroscopy (ARPES) study in $Co_2MnGa$ single crystal shows the position of the nodal lines close to the Fermi level.[28]

The single crystals used in this study are synthesized *via* Bridgman–Stockbarger crystal growth technique. Prior to the transport experiments, the as-grown crystals were characterized by Laue X-ray diffraction (XRD), where sharp spots in the pattern indicate the high crystal quality (see methods for details, Supplementary Fig. S1). The transport experiments were performed in a temperature-variable cryostat (Physical Property Measurement System from Quantum Design) in helium atmosphere, equipped with a 9 T magnet. All transport experiments are performed within the [111]-plane of the $Co_2MnGa$ crystal, with a magnetic field $\mu_0 H$ applied in the [01$\bar{1}$]-direction. In particular, two measurement configurations have been used in our study: First, longitudinal measurements, where the electric voltage response is measured in the direction of the applied electrical current (*I*) or heat current (*Q*) (the corresponding transport parameters are marked with the sub-index xx); and second, transverse measurements, where the electric voltage response is measured normal to the applied electrical current or temperature gradient (the corresponding transport parameters are marked with the sub-index yx, Fig. 1 (d)). To gain deeper insights into the experimental results, *ab-initio* calculations were performed using density-functional theory as implemented in VASP[29] as a first step, followed up by the extraction of a Wannier tight-binding Hamiltonian *via* WANNIER90[30]. With this Hamiltonian, we can evaluate the Berry curvature and consecutively the AHE and ANE (for further details see Methods section).

In the first set of transport experiments, we investigate the longitudinal thermoelectric magneto-transport of $Co_2MnGa$. Fig. 2 (a) shows the electrical resistivity ($\rho_{xx}$) and Seebeck coefficient ($S_{xx}$) as a function of temperature $T$. The decreasing $\rho_{xx}$ with decreasing $T$ reveals the metallic behaviour of the compound with a value of 80 $\mu\Omega$cm at 300 K. Consistently, the magnitude of $S_{xx}$ decreases with lowering of $T$ with a value of around -26 $\mu$V K$^{-1}$ at 300 K. The negative sign of $S_{xx}$ indicates electron dominated transport. The magneto-Seebeck data shows almost no field dependence of $S_{xx}$ within the measured temperature range (Supplementary Fig. S2).

Next, we measure the magnetization ($M$) of our sample. The compound is known to be a ferromagnet with a relatively high Curie temperature of ($T_C$) = 687 K.[31] In Fig. 2 (b) we present temperature-dependent magnetization data for $Co_2MnGa$. In agreement with literature, we find ferromagnetic behavior. From the data taken as a function of sweeping magnetic field, a saturation magnetization value of around $M_{sat}$ = 3.9 $\mu_B$ f.u.$^{-1}$ is measured at 50 K, which slightly decreases with increasing temperature to $M_{sat}$ = 3.7 $\mu_B$ f.u.$^{-1}$ at 300 K.

Further, we measure the anomalous Hall effect of $Co_2MnGa$. We estimate the total Hall conductivity ($\sigma_{xy}$) from the measured Hall resistivity $\rho_{yx}$ and the longitudinal resistivity $\rho_{xx}$ as:

$$\sigma_{xy} = \frac{\rho_{yx}}{\rho_{yx}^2 + \rho_{xx}^2} \qquad (1)$$

In Fig. 2 (c) we present $\sigma_{xy}$ as a function of magnetic field at different base temperatures. As expected for a ferromagnetic material, $\sigma_{xy}$ increases with $\mu_0 H$ and saturates, despite a small slope contributed from the ordinary Hall effect, when the saturation magnetization is reached (compare Fig. 2 (b)). The anomalous Hall conductivity (AHC) ($\sigma_{xy}^A$) has been estimated by extrapolating the slope of the high-field data points (Supplementary Fig. S3) to the zero field value. The observed values of $\sigma_{xy}^A$ and its temperature dependence are consistent with previous reports.[26] As shown in the inset of Fig. 2 (c), a value of $\sigma_{xy}^A$ ~ 1260 S cm$^{-1}$ has been estimated at 60 K, which gradually decreases with increasing temperature and reaches ~845 S cm$^{-1}$ at 300K. From the magnetic field-dependence, one would naively expect $\sigma_{xy}^A$ to only scale with $M$. However, as shown in our *ab-initio* calculation of Fig. 2 (d), the AHC conductivity in Co$_2$MnGa strongly depends on the position of the Fermi level, which is due to the Berry curvature effect discussed above. In the case of the intrinsic AHE, $\Omega_n(k)$ gives an additional term to the group velocity of the electron, which gives:[5]

$$\sigma_{xy}^A = -\frac{e^2}{\hbar} \sum_n \int \frac{dk}{(2\pi)^3} \Omega_{n,xy}(k) f_{nk} \quad (2)$$

where $f_{nk}$ is Fermi-Dirac distribution function with the band index $n$ and the wave vector $k$. $e$ denotes the elementary charge and $\hbar$ is the reduced Planck constant. The integral is performed across the whole Brillouin zone for all bands below $E_F$. Recent results of ARPES measurements on our sample show that the sample is slightly hole-doped.[28] Therefore, the actual Fermi level $E_F$ in our sample is slightly shifted towards the valence bands compared to the theoretical charge neutral point $E_0$ that is estimated by counting the intrinsic electron number by $E_F - E_0 = -80$ meV. Based on the *ab-initio* calculations and ARPES results, we predict an AHC value of $\sigma_{xy}^A(E_F) = 1242$ S cm$^{-1}$ at $T = 0$ K. Extrapolating the measured $\sigma_{xy}^A$ linear to $T = 0$ K, we obtain a reasonably close value of $\sigma_{xy}^A = 1362$ S cm$^{-1}$ which is consistent with the experimentally measured value.

We now turn to investigate the Nernst effect of Co$_2$MnGa. In Fig. 3 (a) the field-dependence of the Nernst thermopower ($S_{yx}$) is shown at different $T$. At all $T$, the Nernst signal increases rapidly at low magnetic fields ($\mu_0 H$) and saturates above $\mu_0 H \sim 1$ T despite a small remaining increase due to ordinary Nernst signal contribution. These features are consistent with the $M$ and $\sigma_{xy}^A$ data. We therefore attribute the saturation value of the Nernst signal to the ANE and the small increasing contribution beyond 1 T to the ordinary Nernst effect. In a similar procedure as for the Hall data, the ordinary and anomalous contributions to the Nernst thermopower can then be separated by fitting the high-field slope of the total Nernst thermopower up to $\mu_0 H = 9$ T (Supplementary Fig. S3). Doing so, we find that the ordinary Nernst effect reaches values of $S_{yx}^O = 50$ nV K$^{-1}$ at room temperature and $\mu_0 H = 1$ T as expected for metallic systems. The ANE, on the other hand, reaches remarkably high values of i.e. $S_{yx}^A$ ~6.0 μV K$^{-1}$ at 300 K, 1 T field (inset of Fig. 3 (a)). Similar to the AHE, the ANE of conventional ferromagnets is proportional to the magnetization, that is, $|S_{yx}^A| = |N_{yx}^A|\mu_0 M$, where $|N_{yx}^A|$ is the anomalous Nernst coefficient, ranging from 0.05 to 1 μV K$^{-1}$ T$^{-1}$ at 300 K. According to this relation, Co$_2$MnGa should maximally exhibit $S_{yx}^A = 0.9$ μV K$^{-1}$ (see methods for details). This value is around 7-times lower value than observed in experiment and indicates that the ANE arises from a mechanism that is distinct from the linear scaling of conventional ferromagnets.

To gain deeper insight into this large ANE in Co$_2$MnGa, we calculate the transverse thermoelectric conductivity ($\alpha_{yx}$) plotted in Fig. 3 (b)). Using the measured components of the $\rho_{xx}$, $\rho_{yx}$, $S_{xx}$ and $S_{yx}$, we estimate $\alpha_{yx}$ via:

$$\alpha_{yx} = \frac{S_{yx}\rho_{xx} - S_{xx}\rho_{yx}}{\rho_{xx}^2 + \rho_{yx}^2} \quad (3)$$

Based on the linear scaling relation for conventional ferromagnetic metals, one would expect that $\alpha_{yx}$ increases linearly with decreasing $T$, following the temperature dependence of $M$.[17] However, in Co$_2$MnGa we observed the inverse effect: $\alpha_{yx}$ increases with increasing temperature although the saturation magnetization decreases. This is another indication that the anomalous Nernst signal in ferromagnetic Co$_2$MnGa arises from a different mechanism than the ANE in conventional ferromagnets. It is noteworthy that the AHC in Co$_2$MnGa also does not follow the conventional scaling relation. This suggests that the ANE and the AHE in Co$_2$MnGa may share a common origin, which is the Berry curvature effect.

To evaluate this possibility, we calculate $\alpha_{yx}$, employing: [5]

$$\alpha_{yx}^{A} = \frac{e}{T\hbar}\sum_{n}\int\frac{dk}{(2\pi)^3}\Omega_{n,yx}(k)\{(\varepsilon_{nk}-\mu)f_{nk}+k_BT\ln[1+e^{-\beta(\varepsilon_{nk}-\mu)}]\} \quad (4)$$

where $\varepsilon_{nk}$ is the band energy. The $\{(\varepsilon_{nk}-\mu)f_{nk}+k_BT\ln[1+e^{-\beta(\varepsilon_{nk}-\mu)}]\}$-term takes finite values only around the Fermi energy. Therefore, while $\sigma_{xy}$ is related to the summation of $\Omega_n(k)$ of the all the occupied bands below $E_F$, $\alpha_{yx}$ is related to $\Omega_n(k)$ of the all bands near $E_F$. Our theoretical calculations show a very large ANC in the system, which is linked to the high Berry curvature induced by the gapping of the nodal lines (Fig. 3 (c)). Like the AHC, the ANC shows a strong Fermi-level dependence, which is not captured by the conventional theory for ferromagnets. Importantly, the calculations reproduce and explain the *T*-dependence of the experimental data as shown in the inset of Fig. 3 (d). At $E_F - E_0 = -80$ meV, the Berry curvature in the model leads to the same characteristic decrease of $\alpha_{yx}^{A}$ with decreasing *T* that is also observed in the experiment. We note that only slight shifts in the position of $E_F$ lead to a strong modification of the *T*-dependence of $\alpha_{yx}^{A}$. Thus, $\alpha_{yx}^{A}(T)$ is a sensitive probe for the underlying mechanism for the ANE in Co$_2$MnGa. The coincident observation of the high values of AHE and ANE, together with the same characteristic *T*-dependence of $\alpha_{yx}^{A}$ in experiment and theory go beyond the conventional model for the ANE in ferromagnets and can be considered as fingerprints of the Berry curvature effect associated with emerging nodal lines and Weyl points.

Finally, we turn to the initial question whether the value of the ANE in such topological ferromagnets can exceed the highest values observed in conventional ferromagnets. Therefore, we compare in Fig. 4 the anomalous Nernst signal of Co$_2$MnGa with the values of various other ferromagnetic metals,[9,20,32–36] extending the plot of Ikhlas *et al.*[9] We plot the absolute values of the Nernst thermopower in the magnetically ordered states, where $\mu_0H$ and *T* are kept as absolute parameters. The shaded area of the graph represents the linear relation of the anomalous Nernst thermopower as measured for conventional ferromagnets. The red dotted line marks the limit of the highest value of $|S_{yx}^{A}|$ ever measured in a ferromagnet. As explained above, according to the conventional scaling relation the expected value of $S_{yx}^A$ for Co$_2$MnGa at room temperature is ~0.05 to 0.9 µV K$^{-1}$. Nevertheless, the Nernst signal of Co$_2$MnGa shows a much larger value ~ 6.0 µV K$^{-1}$ at room temperature and further increases to ~6.6 µV K$^{-1}$ at 340 K, the highest temperature investigated.

In summary, we have observed a remarkably large ANE in the ferromagnetic topological Heusler compound Co$_2$MnGa which is around 7-times larger than any value reported for conventional ferromagnets so far in the literature. Combined electrical, thermoelectric and first-principles calculations reveal that this high value of the ANE arises from a large net Berry curvature near the Fermi energy associated with nodal lines and Weyl points. While our results not only show that the ANE in ferromagnets originates generally from a net Berry curvature of all bands near $E_F$ rather than directly from the magnetization, the large Nernst signal at room temperature and high Curie temperature also make Co$_2$MnGa an ideal candidate for further investigation for use in transverse thermoelectric device. In a broader sense, the flexible electronic structure of Heusler compounds can open up adequate room for the realization of a giant Nernst effect *via* a suitable Berry curvature design. We believe, the present study can act as a guide in search for new magnetic topologically non-trivial materials with large non-zero Berry curvature for observation of a giant anomalous Nernst effect.

# REFERENCES

1. Rowe, D. M. *CRC Handbook of Thermoelectrics. CRC Handbook of Thermoelectrics* (CRC Press, Boca Raton, FL, 1995).
2. Nolas, G. S., Sharp, J. & Goldsmid, J. *Thermoelectrics: basic principles and new materials developments.* **45,** (Springer Science & Business Media, 2013).
3. Nagaosa, N., Sinova, J., Onoda, S., MacDonald, A. H. & Ong, N. P. Anomalous Hall effect. *Rev. Mod. Phys.* **82,** 1539–1592 (2010).
4. Pu, Y., Chiba, D., Matsukura, F., Ohno, H. & Shi, J. Mott relation for anomalous Hall and Nernst effects in Ga1-xMnxAs ferromagnetic semiconductors. *Phys. Rev. Lett.* **101,** 117208 (2008).
5. Xiao, D., Yao, Y., Fang, Z. & Niu, Q. Berry-phase effect in anomalous thermoelectric transport. *Phys. Rev. Lett.* **97,** 26603 (2006).
6. Xiao, D., Chang, M. C. & Niu, Q. Berry phase effects on electronic properties. *Rev. Mod. Phys.* **82,** 1959–2007 (2010).
7. Liang, T. *et al.* Anomalous Nernst Effect in the Dirac Semimetal Cd3As2. *Phys. Rev. Lett.* **118,** 136601 (2017).
8. Jia, Z. *et al.* Thermoelectric signature of the chiral anomaly in Cd3As2. *Nat. Commun.* **7,** 13013 (2016).
9. Ikhlas, M. *et al.* Large anomalous Nernst effect at room temperature in a chiral antiferromagnet. *Nat. Phys.* **13,** 1085–1090 (2017).
10. Thierschmann, H. *et al.* Three-terminal energy harvester with coupled quantum dots. *Nat. Nanotechnol.* **10,** 854–858 (2015).
11. Bergenfeldt, C., Samuelsson, P., Sothmann, B., Flindt, C. & Büttiker, M. Hybrid microwave-cavity heat engine. *Phys. Rev. Lett.* **112,** 76803 (2014).
12. Sánchez, D. & Serra, L. Thermoelectric transport of mesoscopic conductors coupled to voltage and thermal probes. *Phys. Rev. B* **84,** 201307 (2011).
13. Entin-Wohlman, O., Imry, Y. & Aharony, A. Three-terminal thermoelectric transport through a molecular junction. *Phys. Rev. B* **82,** 115314 (2010).
14. Behnia, K. & Aubin, H. Nernst effect in metals and superconductors: a review of concepts and experiments. *Rep. Prog. Phys.* **79,** 46502 (2016).
15. Behnia, K. The Nernst effect and the boundaries of the Fermi liquid picture. *J. Phys. Condens. Matter* **21,** 113101 (2009).
16. Watzman, S. J. *et al.* Dirac dispersion generates large Nernst effect in Weyl

semimetals. *Phys. Rev. B* **97,** 161404 (2018).

17. Miyasato, T. *et al.* Crossover behavior of the anomalous Hall effect and anomalous nernst effect in itinerant ferromagnets. *Phys. Rev. Lett.* **99,** 86602 (2007).

18. Lee, W. L., Watauchi, S., Miller, V. L., Cava, R. J. & Ong, N. P. Anomalous Hall heat current and Nernst effect in the $CuCr_2Se_{4-x}Br_x$ ferromagnet. *Phys. Rev. Lett.* **93,** 226601 (2004).

19. Smith, W. A. The transverse thermomagnetic effect in nickel and cobalt. *Phys. Rev.* **33,** 295–306 (1911).

20. Ramos, R. *et al.* Anomalous Nernst effect of $Fe_3O_4$ single crystal. *Phys. Rev. B* **90,** 54422 (2014).

21. Burkov, A. A. Anomalous Hall effect in Weyl metals. *Phys. Rev. Lett.* **113,** 187202 (2014).

22. Nayak, A. K. *et al.* Large anomalous Hall effect driven by a nonvanishing Berry curvature in the noncolinear antiferromagnet $Mn_3Ge$. *Sci. Adv.* **2,** e1501870–e1501870 (2016).

23. Nakatsuji, S., Kiyohara, N. & Higo, T. Large anomalous Hall effect in a non-collinear antiferromagnet at room temperature. *Nature* **527,** 212–215 (2015).

24. Li, X. *et al.* Anomalous Nernst and Righi-Leduc Effects in $Mn_3Sn$: Berry Curvature and Entropy Flow. *Phys. Rev. Lett.* **119,** 56601 (2017).

25. Kiyohara, N., Tomita, T. & Nakatsuji, S. Giant Anomalous Hall Effect in the Chiral Antiferromagnet $Mn_3Ge$. *Phys. Rev. Appl.* **5,** 64009 (2016).

26. Manna, K. *et al.* From colossal to zero: Controlling the Anomalous Hall Effect in Magnetic Heusler Compounds via Berry Curvature Design. *arXiv:1712.10174* (2017).

27. Kübler, J. & Felser, C. Weyl points in the ferromagnetic Heusler compound $Co_2MnAl$. *EPL* **114,** 47005 (2016).

28. Belopolski, I. *et al.* A three-dimensional magnetic topological phase. *arXiv:1712.09992* (2017).

29. Kresse, G. & Furthmüller, J. Efficient iterative schemes for ab initio total-energy calculations using a plane-wave basis set. *Phys. Rev. B* **54,** 11169–11186 (1996).

30. Mostofi, A. A. *et al.* wannier90: A tool for obtaining maximally-localised Wannier functions. *Comput. Phys. Commun.* **178,** 685–699 (2008).

31. Ido, H. & Yasuda, S. Magnetic properties of Co-Heusler and related mixed alloys. *Le J. Phys. Colloq.* **49,** C8-141 (1988).

32. Hasegawa, K. *et al.* Material dependence of anomalous Nernst effect in perpendicularly magnetized ordered-alloy thin films. *Appl. Phys. Lett.* **106,** 252405 (2015).

33. Hanasaki, N. *et al.* Anomalous Nernst effects in pyrochlore molybdates with spin chirality. *Phys. Rev. Lett.* **100,** 106601 (2008).

34. Weischenberg, J., Freimuth, F., Blügel, S. & Mokrousov, Y. Scattering-independent anomalous Nernst effect in ferromagnets. *Phys. Rev. B* **87,** 60406 (2013).

35. Shiomi, Y., Kanazawa, N., Shibata, K., Onose, Y. & Tokura, Y. Topological Nernst effect in a three-dimensional skyrmion-lattice phase. *Phys. Rev. B* **88,** 64409 (2013).

36. Uchida, K. I. *et al.* Enhancement of anomalous Nernst effects in metallic multilayers free from proximity-induced magnetism. *Phys. Rev. B* **92,** 94414 (2015).

# FIGURES

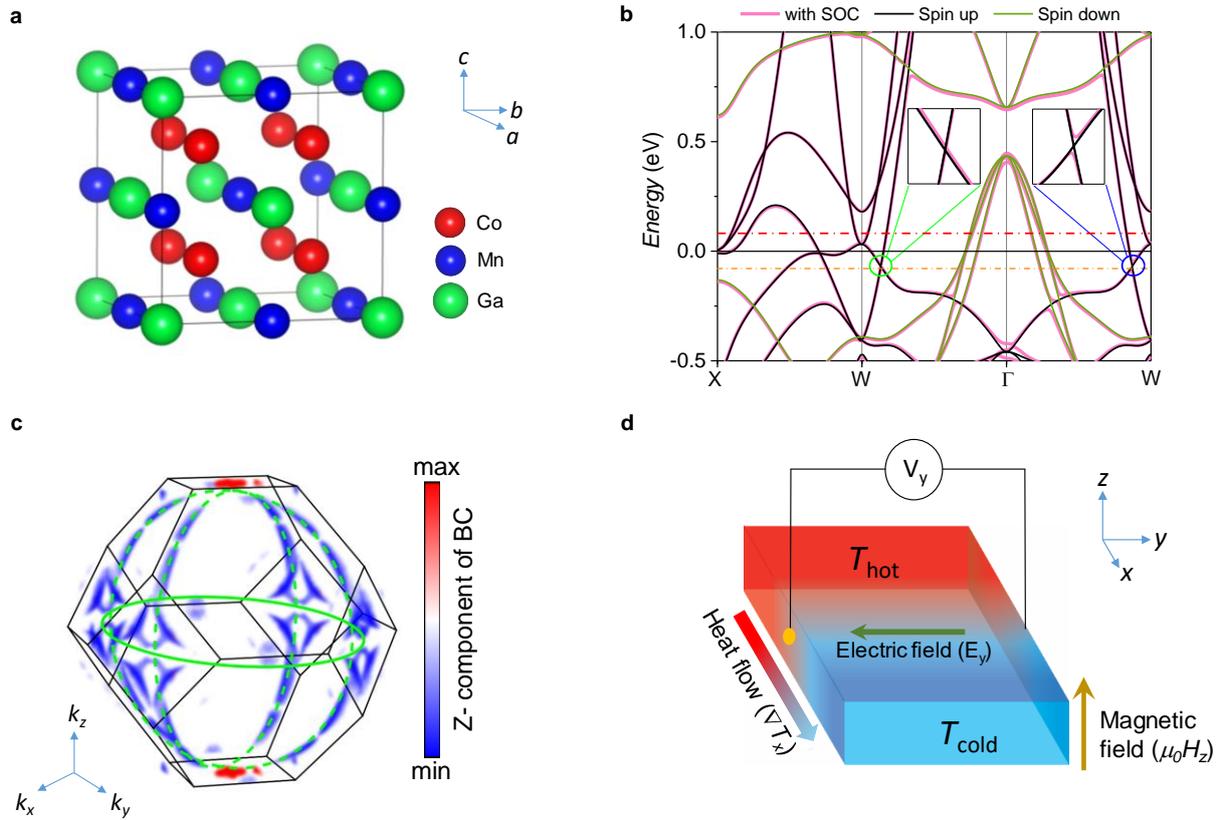

**Figure 1 | Crystal and electronic structure and Nernst measurement geometry. a**, Crystal structure of $Co_2MnGa$. The structure consists of four interpenetrating fcc sub-lattices. Two of the sublattices are formed by Co atoms and the other two by the Mn and Ga atoms. **b**, Electronic structure of $Co_2MnGa$ with and without spin-orbit coupling (SOC). The nodal line regions are marked with circles and zoomed in the inset. SOC induces a gapping in the nodal line in the blue colored circle by lifting the degeneracy of the bands. The dashed lines indicate different positions of the Fermi level ($E_F$) with respect to the charge neutrality point $E_0$. **c**, Berry curvature distribution in Brillouin zone. The full and dotted green lines indicate the nodal lines. The Berry curvature is concentrated around the gapped lines (dashed). **d**, Schematic representation of the Nernst measurement. In this geometry, a transverse thermoelectric voltage $V_y$ is produced by a temperature gradient $\nabla T_x$ and a magnetic field $\mu_0 H$ normal to each other.

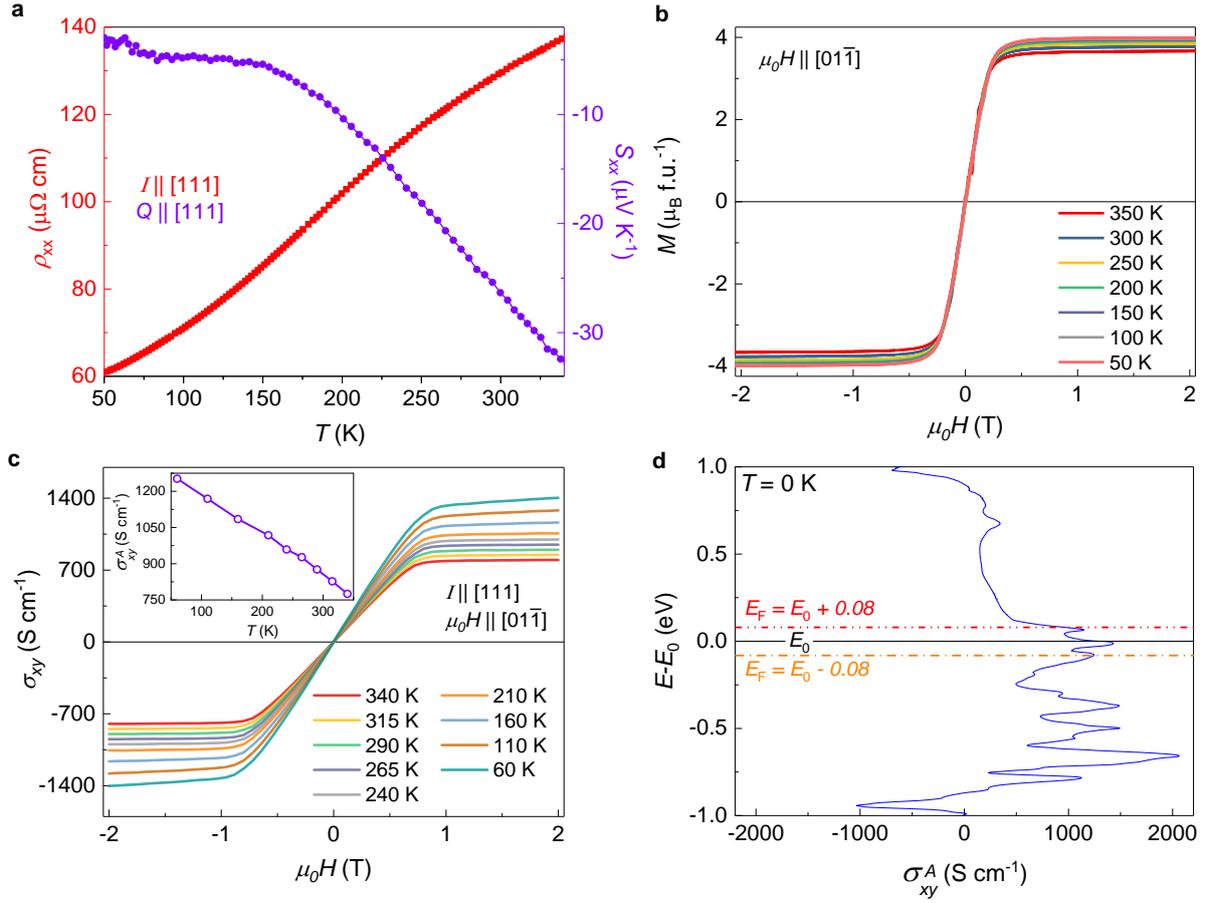

**Figure 2 | $T$-dependent resistivity and Seebeck coefficient, magnetization and Hall conductivity of $Co_2MnGa$. a**, Temperature ($T$)-dependent electrical resistivity ($\rho_{xx}$) and zero-field Seebeck coefficient ($S_{xx}$). Electrical ($I$) and heat current ($Q$) has been applied along [111] direction of the crystal **b**, Magnetic field-dependence of the magnetization ($M$) at different $T$. **c**, Magnetic field-dependence of the Hall conductivity ($\sigma_{xy}$). The Hall conductivity has been extracted employing the measured off-diagonal ($\rho_{yx}$) and diagonal components ($\rho_{xx}$) of the resistivity tensor. The inset shows temperature dependence of the anomalous Hall conductivity ($\sigma_{xy}^A$). **d**, Theoretically estimated Hall conductivity as a function of chemical potential at $T = 0$ K for $Co_2MnGa$. The dashed lines indicate different positions of the Fermi level ($E_F$) with respect to the charge neutrality point $E_0$.

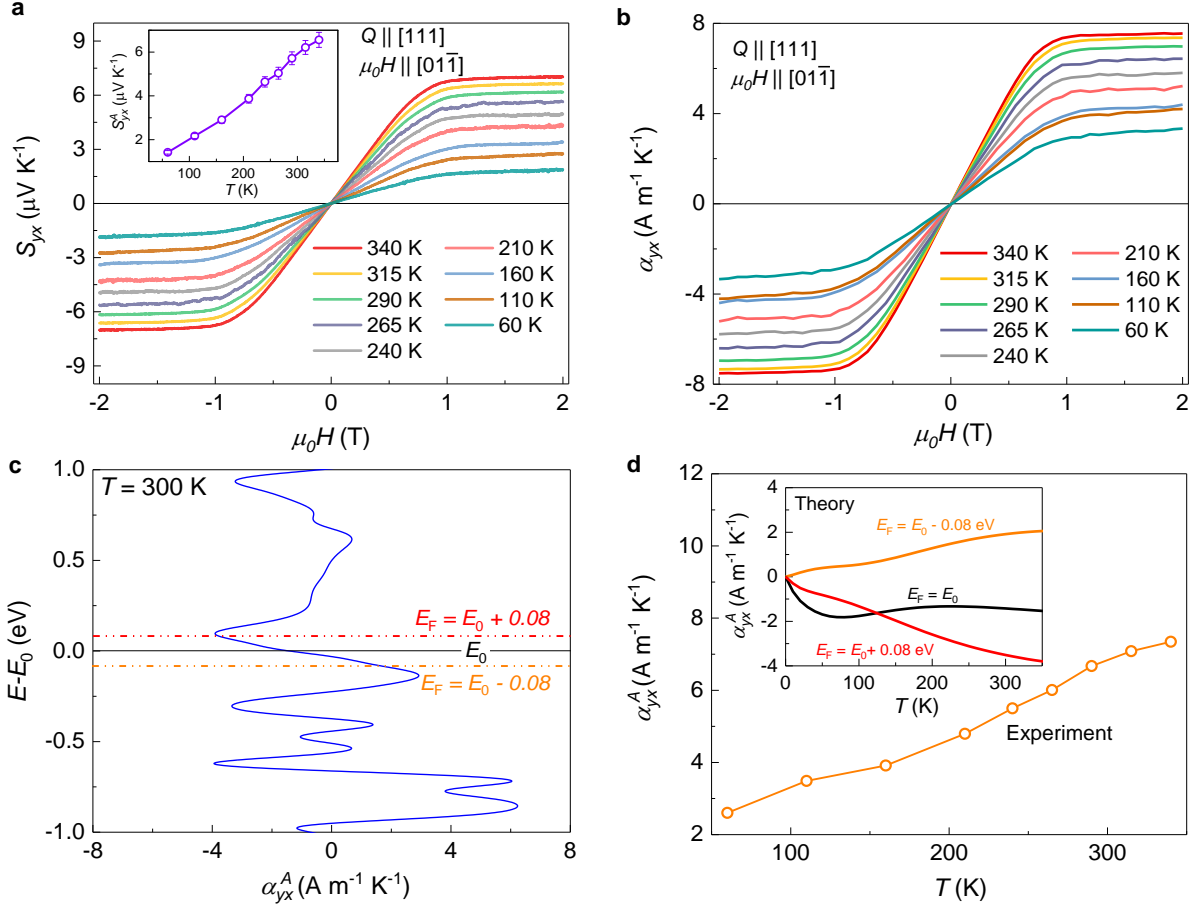

**Figure 3 | Nernst effect of $Co_2MnGa$. a,** Magnetic field-dependence of the Nernst thermopower ($S_{yx}$) of $Co_2MnGa$ at different temperatures *T*. The inset shows the extracted anomalous Nernst thermopower ($S_{yx}^A$) extrapolating the slope of the high-field (above 1 T) data points as a function of *T*. **b,** Magnetic field-dependence of the transverse thermoelectric conductivity ($\alpha_{yx}$) at different *T*. **c,** Theoretically calculated anomalous Nernst conductivity ($\alpha_{yx}^A$) as a function of the chemical potential with respect to the charge neutrality point $E_0$ at *T* = 300 K for $Co_2MnGa$. The dashed lines indicate different position of the Fermi level $E_F$. **d,** *T*-dependence of the experimentally determined $\alpha_{yx}^A$. The inset of the plot shows the theoretically estimated temperature dependence of $\alpha_{yx}^A$ at the Fermi level positions marked in (c).

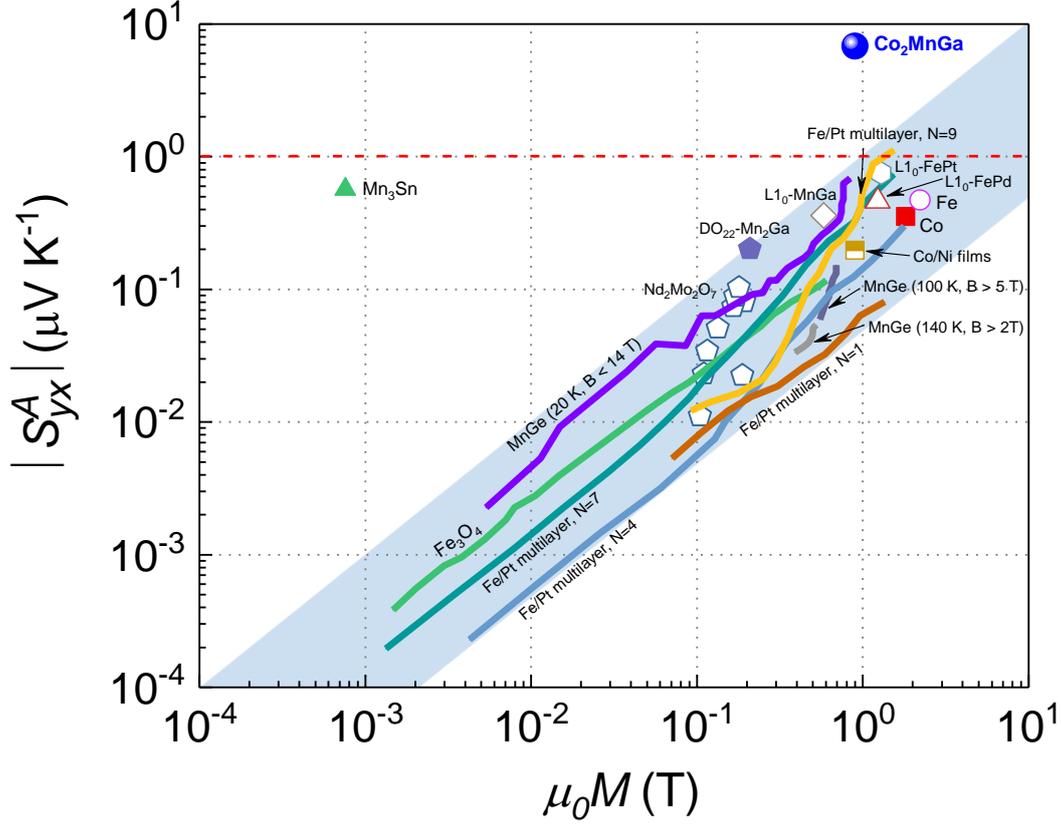

**Figure 4 | Magnetization-dependence of the anomalous Nernst thermopower.** Comparison of the magnetization-dependent anomalous Nernst thermopower ($|S_{yx}^A|$) of various ferromagnetic metals (see methods for details), chiral antiferromagnet Mn$_3$Sn and Co$_2$MnGa extending the plot of Ikhlas *et al.*[9] The shaded region indicates the linear relation for conventional ferromagnetic metals $|S_{yx}^A| = |N_{yx}^A|\mu_0 M$, with $|N_{yx}^A|$ ranging from ~0.05 µV K$^{-1}$ T$^{-1}$ to ~1 µV K$^{-1}$ T$^{-1}$. The red dotted line indicates the upper limit of the measured value of $|S_{yx}^A|$ in a conventional ferromagnet. The Nernst signal for Co$_2$MnGa falls out of the scope of this relation, similar as Mn$_3$Sn. The $|S_{yx}^A|$ for Co$_2$MnGa shows the largest value reported among ferromagnetic metals at room temperature.

# METHODS

### Co$_2$MnGa single-crystal growth

Single crystals of Co$_2$MnGa were grown using the Bridgman Stockbarger crystal growth technique. For growing a single crystal, first, we have synthesized a polycrystalline sample by induction melting from the high-quality stoichiometric amounts of elemental cobalt (99.999%), manganese (99.999%) and gallium (99.999%) in an alumina crucible. The induction-melted sample is then crushed into powder form and packed in a custom-designed sharp-edged alumina tube, which was sealed in a tantalum tube. The sample was heated to 1523 K and soaked for 10 h to ensure the homogeneity of the melt, and then slowly cooled to 1023 K. Single crystallinity of the as-grown crystal was checked by white-beam backscattering Laue X-ray diffraction at room temperature. The aligned crystal was cut into bar-shaped samples for transport and magnetization measurements. The typical dimension of the crystals used for electrical and thermal transport measurements is $7.5 \times 2 \times 0.5$ mm$^3$.

### Magnetization and electrical transport measurement

Magnetization measurement was performed using a Quantum Design vibrating sample magnetometer (MPMS). The electrical transport properties were measured in a Quantum Design physical property measurement system (PPMS, ACT option). The standard four-probe method has been used for electrical transport measurement. To correct for contact misalignment, the measured raw data were field symmetrized and antisymmetrized, respectively.

### Thermoelectric transport measurements

The zero-field temperature dependent Seebeck coefficient was measured by the one-heater and two-thermometer configuration using the thermal transport option (TTO) of the PPMS (Quantum Design). Magneto-Seebeck and Nernst thermoelectric measurement were carried out by the one-heater and two-thermometer configuration on a PPMS. The instrument was controlled by software programmed using LabVIEW. The measurement was done in the temperature range 60- 340 K, and magnetic fields were swept in both directions. The temperature gradient has been generated using a resistive heater, connected to the gold-coated copper plate at one end of the sample. The thermal gradient, $\nabla T$ was applied along the [111] direction of the sample and magnetic field was applied along [01$\bar{1}$] direction. For the heat sink, a gold-plated copper plate was attached to the puck clamp. To measure the temperature difference, two gold-plated copper leads were attached directly to the sample using silver epoxy along the thermal gradient direction. The distance between the thermometers is ~4 mm. The $\nabla T$ was typically set to be 1 % - 3 % of the sample temperature. The transverse voltage due to the thermal gradient was measured by attaching two copper wires, orthogonal to the heat gradient direction of the sample using silver epoxy. The Seebeck thermopower was estimated using the relation, $S_{xx} = E_x/\nabla T_x$, where $E_x$ is longitudinal electric field. The Nernst thermopower was estimated as $S_{yx} = L_x E_y/(L_y \nabla T_x)$, where $E_y$ is transverse electric field, $L_x$ is the distance between two temperature leads, and $L_y$ is the distance between two voltage wires.

**Nernst data survey**

We reproduce the magnetization dependent anomalous Nernst thermopower value for different ferromagnets well below their Curie temperatures from the previous report by Ikhlas et al.[9] $Mn_3Sn$ (200 K; ref. 9 ), $Fe_3O_4$ (300 K, B < 0.8 T; ref. 20), Co/Ni films (300 K; ref. 32), $L1_0$ –FePt (300 K; ref. 32), $D0_{22}$ -$Mn_2Ga$ (300 K; ref. 32), $L1_0$ -MnGa (300 K; ref. 32), $L1_0$ -FePd (300 K; ref. 32), $Nd_2Mo_2O_7$ (T < Tc = 73 K, B = 1 T [111]; ref. 33), Fe (300 K; ref. 34), Co (300 K; ref. 34), MnGe (140 K, B > 2 T; ref. 35), MnGe (100 K, B > 5 T; ref. 35), MnGe (20 K, B < 14 T; ref. 35), and Pt/Fe multilayer N = 1 ~ 9 (300 K, B < 5 T; ref. 36).

### *ab-initio* calculations

For our *ab-initio* calculations, we employ density-functional theory (DFT) using VASP[29], describing the exchange-correlation energy via PBE. For integrations in k-space we use a grid of 19 x 19 x 19 points. We extract Wannier functions from the resulting band structure using WANNIER90[30] to set up a tight-binding Hamiltonian, which reproduces the DFT band structure within a few meV. With this Hamiltonian we calculate the Berry curvature $\vec{\Omega}_n$ via

$$\Omega_{n,ij} = \sum_{m \neq n} \frac{\langle n|\frac{\partial H}{\partial k_i}|m\rangle \langle m|\frac{\partial H}{\partial k_j}|n\rangle - (i \leftrightarrow j)}{(\varepsilon_n - \varepsilon_m)^2},$$

where *m* and *n* are the eigenstates and $\varepsilon$ are the eigenenergies of the Hamiltonian H. From this we can calculate the anomalous Hall conductivity $\sigma_{xy}^A$ using

$$\sigma_{xy}^A = -\frac{e^2}{\hbar} \sum_n \int \frac{dk}{(2\pi)^3} \Omega_{n,xy}(k) f_{nk}$$

with $f_{nk}$ as the Fermi-Dirac distribution for a band *n* at a k-point *k* and the anomalous Nernst conductivity $\alpha_{yx}^A$ as described in (ref. 5)

$$\alpha_{yx}^A = \frac{e}{T\hbar} \sum_n \int \frac{dk}{(2\pi)^3} \Omega_{n,yx}(k) \{(\varepsilon_{nk} - \mu) f_{nk} + k_B T \ln[1 + e^{-\beta(\varepsilon_{nk} - \mu)}]\}$$

with the Boltzmann constant $k_B$ and $\beta = k_B T$.

To realize integrations over the whole Brillouin zone for AHE and ANE we employ a mesh of $251 \times 251 \times 251$ k-points which was carefully checked to be converged.

### ACKNOWLEDGEMENTS

This work was financially supported by the ERC Advanced Grant No. (742068) "TOP-MAT". Ch.F. acknowledges financial support from the Alexander von Humboldt Foundation.

# AUTHOR INFORMATION

## Contributions

S.N.G. carried out thermoelectric transport measurements with the help of S.J.W. and Ch.F. K.M. synthesized the single-crystal and carrier out electrical transport measurement. J.N. carried out the *ab-initio* calculations with the help of Y.S.. S.N.G., S.J.W., K.M., J.N., Y.S. and J.G. analyzed the data. J.G. and C.F. supervised the project. All authors contributed to the writing of the manuscript.

## Data availability statement

All data generated or analysed during this study are included in this published article (and its supplementary information). The datasets generated during and/or analysed during the current study are available from the corresponding author on reasonable.

## Competing financial interest

The authors declare no competing financial interests.

## Corresponding author

*satyanarayanguin@cpfs.mpg.de, Johannes.Gooth@cpfs.mpg.de, Claudia.Felser@cpfs.mpg.de*

# Supporting Information

# Anomalous Nernst effect beyond the magnetization scaling relation in the ferromagnetic Heusler compound $Co_2MnGa$

Satya N. Guin[1,*], Kaustuv Manna[1], Jonathan Noky[1], Sarah J. Watzman[2], Chenguang Fu[1], Nitesh Kumar[1], Walter Schnelle[1], Chandra Shekhar[1], Yan Sun[1], Johannes Gooth[1,*], Claudia Felser[1,*]

[1] Max Planck Institute for Chemical Physics of Solids, 01187 Dresden, Germany

[2] Department of Mechanical and Aerospace Engineering, The Ohio State University, Columbus, OH 43210, USA

*Correspondence to: satyanarayanguin@cpfs.mpg.de

Johannes.Gooth@cpfs.mpg.de

Claudia.Felser@cpfs.mpg.de

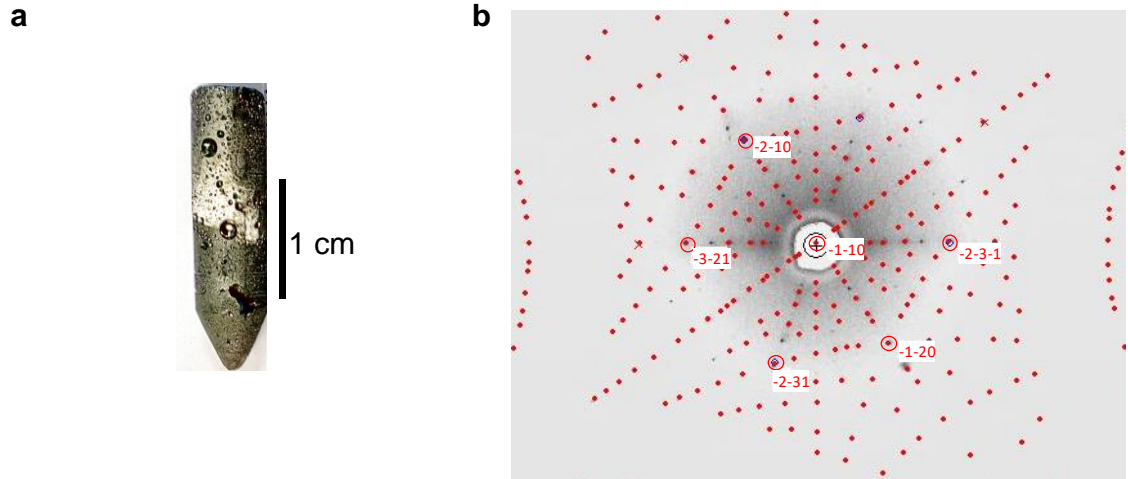

**Figure S1 | Picture of as grown crystal and Laue diffraction.** (a) Bridgman Stockbarger method grown single crystal of Co$_2$MnGa. The single crystal nature of the sample has been confirmed using Laue x-ray diffraction technique measurement. The diffraction pattern of the as crystal can be indexed based on $Fm\bar{3}m$ space group and a lattice constant $a = 5.77$ Å (b) Laue diffraction pattern of [-1-10] direction oriented crystal superposed with a theoretically simulated pattern. The oriented crystal was then cut using wire saw into bar-shaped for transport and magnetization measurements.

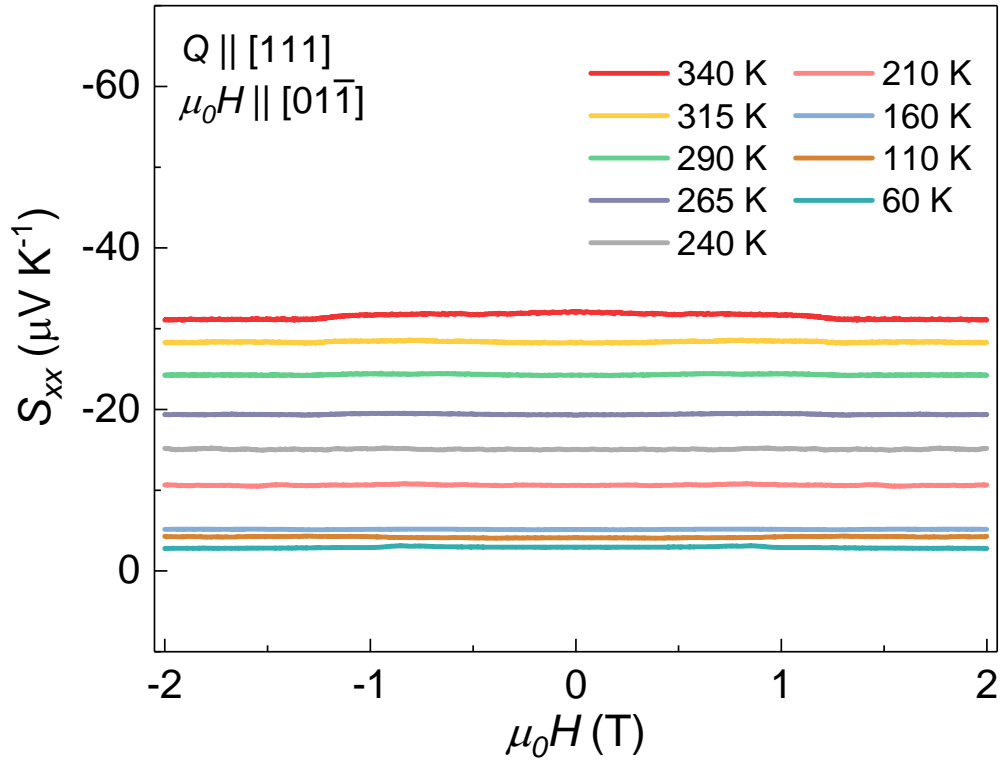

**Figure S2 | Magneto-Seebeck thermopower ($S_{xx}$) for Co$_2$MnGa.** $S_{xx}$ show almost no field dependency in all the measurement temperature.

In Co$_2$MnGa, the zero field Seebeck thermopower ($S_{xx}$) value decreases with lowering the base temperature of measurement. The is due to metallic nature of the sample as evident from the longitudinal resistivity data ($\rho_{xx}$).

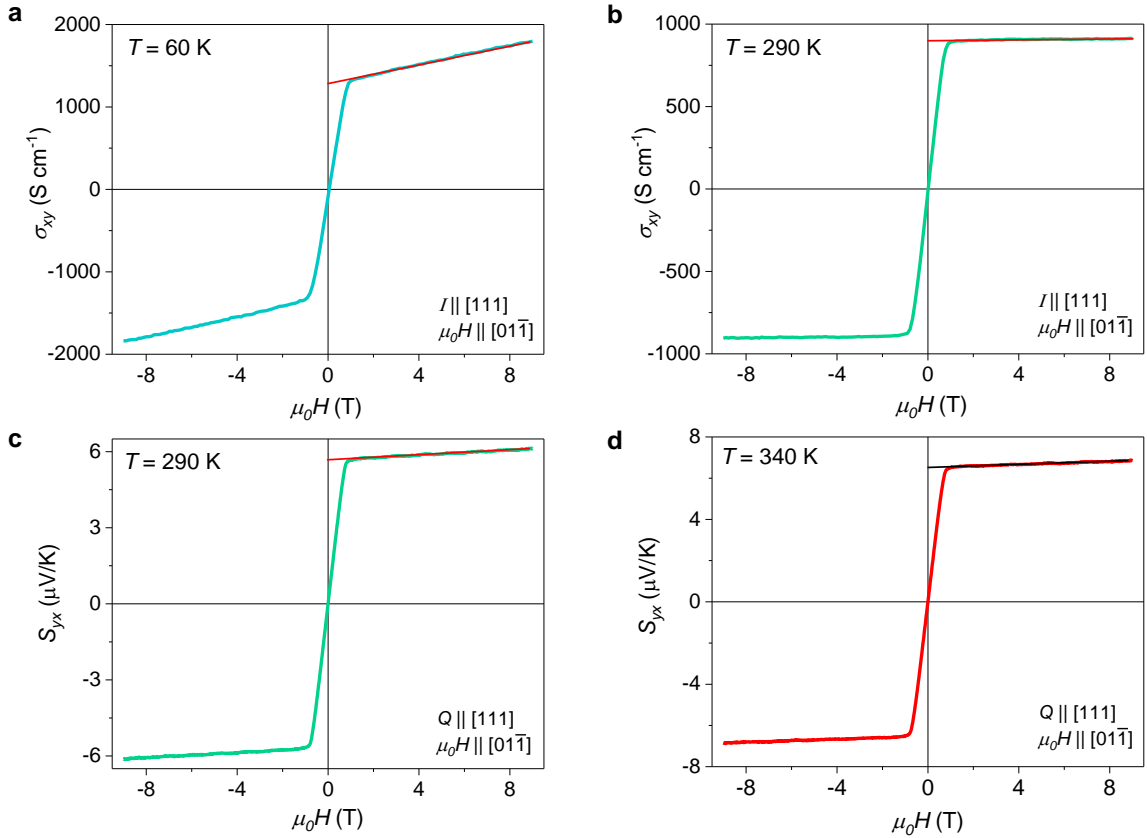

**Figure S3 | High field Hall and Nernst data.** Hall conductivity ($\sigma_{xy}$) and Nernst thermopower ($S_{yx}$) of Co$_2$MnGa at selected temperature up to 9T magnetic field. The high field data has been used for liner fitting. The anomalous response has been estimated by extrapolation of slope to the zero field value.

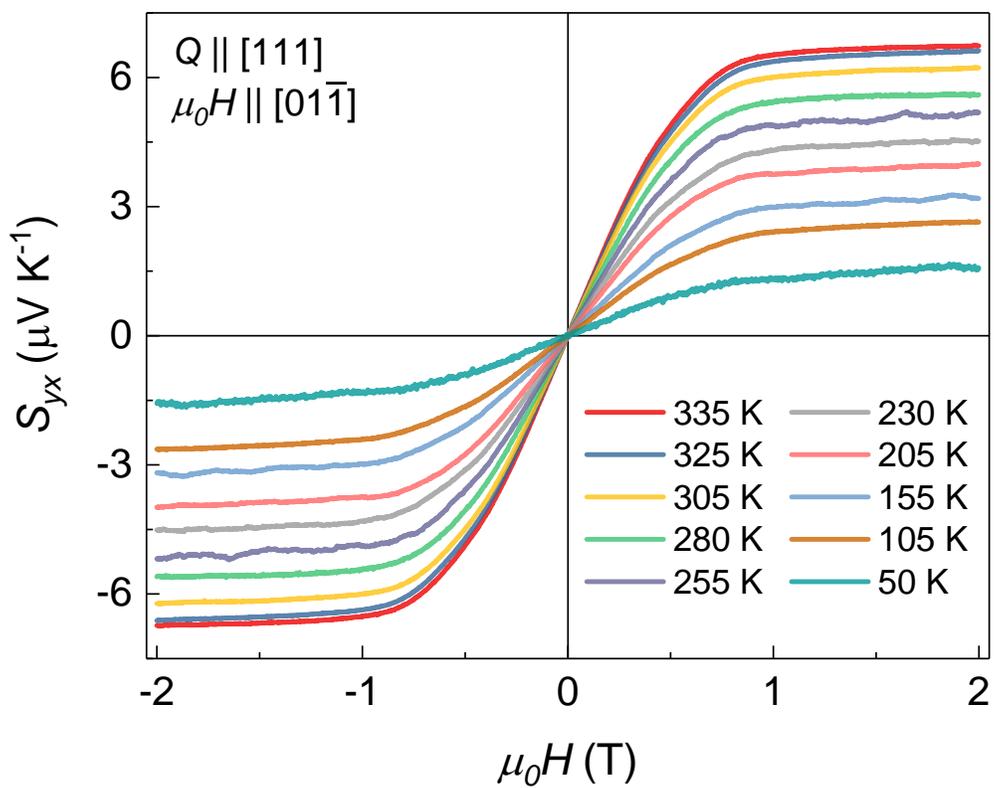

**Figure S4 | Nernst thermopower on a different Co$_2$MnGa sample.** The Nernst thermopower ($S_{yx}$) data shows a similar result indicating the reproducibility of the observed value.